# Anomalous Redshift of Some Galactic Objects

Yi-Jia Zheng

National Astronomical Observatory, Chinese Academy of Sciences, Beijing, 100012 China

**Abstract**

Anomalous redshifts of some galactic objects such as binary stars, early-type stars in the solar neighborhood, and *O* stars in a star clusters are discussed. It is shown that all these phenomena have a common characteristic, that is, the redshifts of stars increase as the temperature rises. This characteristic cannot be explained by means of the Doppler Effect but can by means of the soft-photon process proposed by Yijia Zheng (arXiv:1305.0427 [astro-ph.HE]).

**Keywords:** redshift, binary stars, early-type stars, star clusters, k effect

## I. Introduction

Many observed redshifts of galactic objects cannot be explained by means of the ordinary Doppler Effect interpretation. For example, for the redshifts of individual spectrum of each component of many binary stars, the ordinary Doppler Effect can explain the variation part, but not the difference between the average redshifts of the two components. The observations of stars close to the sun show that the early-type stars have a large positive recessing velocity (k effect). The apparent recession of early-type stars in the solar neighborhood appears suspicious if one gives them a Doppler Effect interpretation. The receding velocity of *O* stars relative to the average star velocity in the same cluster was also observed. All these phenomena have a common characteristic, that is, the higher the temperature of a star is, and the larger the observed redshift is. According to the model atmosphere of stars, the electron number density of stars increases as the temperature of the stars rises (Kurucz 1979). This means that the fact that the redshifts of stars increase with the temperature is related to the increase of number of electrons along the sight line. This is a strong support to the soft-photon process theory proposed by Yijia Zheng (arXiv:1305.0427 [astro-ph.HE]).

## II. Examples

1. Binary stars

The variation of wavelength of each component of the binary system is a function of time. That time-dependent variation is a Doppler effect due to the velocity of rotation of the two components around their center of mass. For example, an oscillatory variation of spectral line due to the radial velocity is shown in fig 1.

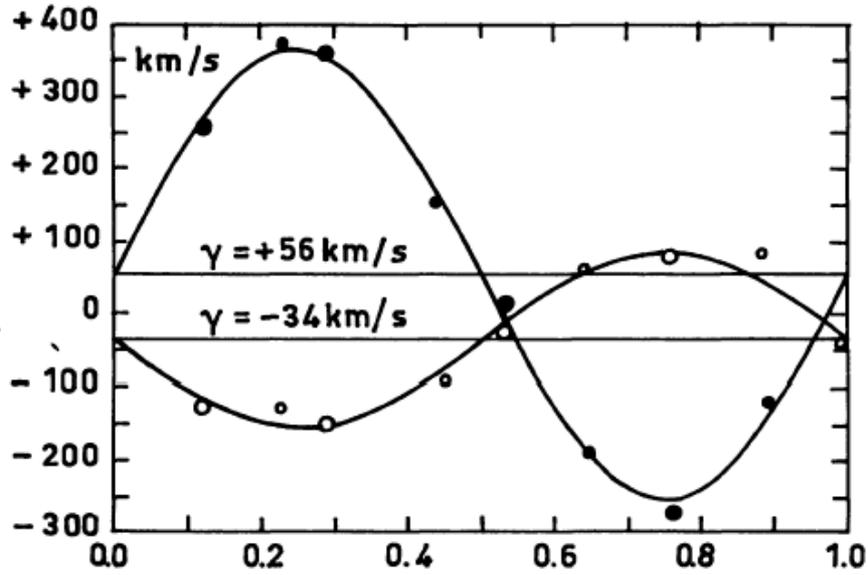

Fig. 1. Velocity curves for HD 193567 (Wilson, 1940). The curve with γ=+56 km/s is obtained from the emission line He II 4686; the other from the absorption lines of the companion (from Knhi 1974).

In the figure one can see the observed redshift of each star at different phases of rotation. The inverted phase observed between components easily proves the binary nature of the system. However, a problem arises because it is observed that the average radial velocity of each component of the system is not the same, but celestial mechanics predicts that when two bodies rotate around their center of mass, the average radial velocity of either star with respect to us must be the same. This problem in many cases can be seen in table 1 and fig. 1.

**Table 1** Difference of velocities between the centers of mass of binary components of some systems (reproduced from Marmet 1990).

| HD | Δv (km/s) | Spectral type |
|---|---|---|
| 68273 | 82 | WC7+O91 |
| 94546 | 78 | WN4+OB |
| 152248 | 52 | |
| 168206 | 190 | WCB+BO |
| 186943 | 105 | WN4+B |
| 190918 | 126.3 | WN4+O9 |
| 191765 | 185 | W-R |
| 192103 | 280 | W-R |
| 192163 | 205 | W-R |
| 193576 | 90 | W-R |
| 193928 | 93 | WN6+BOI |
| 211853 | 105 | WN6+BOI |
| 228766 | 90 | WN7+O |

It can be seen that the component with a much higher temperature has a larger redshift than the other one. Wilson (1949) concluded: "Whatever the origin of the shift may be, it is probably to be sought in the process occurring within the star atmosphere itself." Cawley and Hutchings (1974) have tried to eliminate the velocity difference ∆v by reexamining the original data, but it is found that difficulties of measuring blends and asymmetrical profiles can account for a large fraction of the velocity differences, but the differences in the two components still present.

Marmet (1988) has demonstrated that the electromagnetic interaction on atoms is inelastic and leads to a non-Doppler redshift. But Marmet did not show that the number of atoms of the stars increased with the temperature of the stars. Therefore, even the electromagnetic interaction on atoms is inelastic and leads to a non-Doppler redshift, this effect cannot explain that the redshift of a star increases with the temperature. According to the atmosphere model of stars proposed by Kurucz (1979), the electron number densities of stars increase with the temperature of the stars. This means that the electron density $n_e$ is larger in the atmosphere of the star with a higher temperature. Due to the soft-photon process it will lead to a larger redshift when the radiation originates from a hotter star. This is a strong support to the theory proposed by Yijia Zheng in arXiv:1305.0427 [astro-ph.HE].

2. K effect

Radial velocity of 2148 stars close to the sun have been analyzed by Cambell and Moore (1928) and by Smart and Green (1936). Torondjadze (1953) thought that the dynamic properties of the later-type stars are different from those of the young stars. The early-type stars have their own large positive recessing velocity ($K$ term) as if they originated from a location near the Sun and formed an expanding association arranged along the spiral arm of the Milky Way. The apparent recession of early-type stars in the solar neighborhood appears suspicious if one gives them a Doppler interpretation. Marmet (1990) found that the apparent recession of stars depends on their temperature. Fig.2 shows the relationship between the recession velocity and the temperature (from Marmet 1990). This is of compatibility with the non-Doppler interpretation that the redshift occurring within the star atmosphere is due to the soft photon emitting during the light propagation.

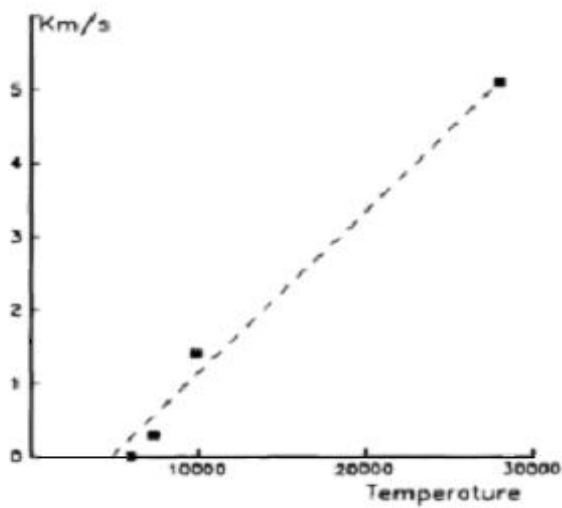

Fig.2. Apparent velocity of recession in all directions for stars in the solar neighborhood as a function of the temperature of the star (from Marmet 1990).

3. Redshifts in star clusters

The most conspicuous young star cluster in our own Milk Way Galaxy is double cluster h+$x$ Persei. This is the association called Per OB 1 by Humphreys (1978) who listed radial velocity for 84 members. Fig. 3 shows that the most luminous stars in this association have the highest redshifts. They progressively decrease to about 15 km s$^{-1}$ smaller at 2.5 mag fainter luminosity. Olano and Poppel recognized that the velocity of early-type stars is larger than that of the surrounding stars. The receding velocity of $O$ stars relative to the average star velocity in the same cluster was first observed by Trumpler (1935). He attempted to use this redshift to determine the mass of $O$ stars, but it was discovered later that most of this redshift could not have a relativistic origin. The general redshift of "runaway stars" observed can only be explained by the non-Doppler interpretation, since the redshift increases as a function of the temperature of the star.

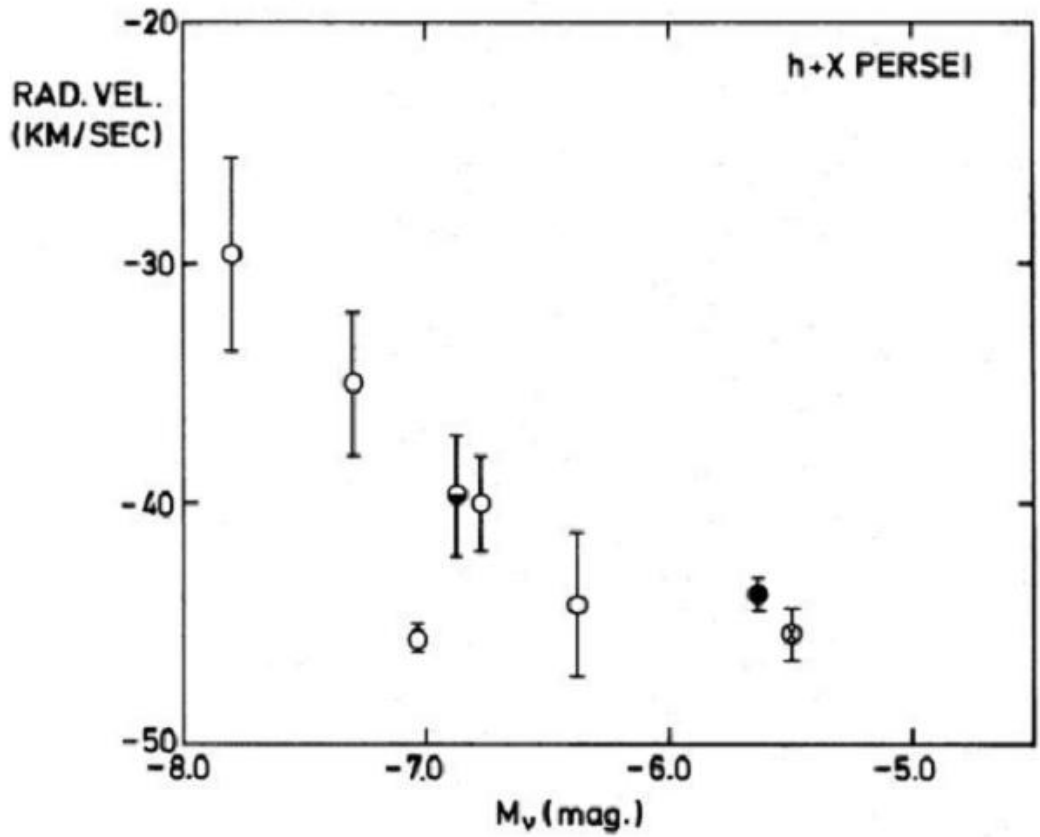

Fig.3. Redshift of most luminous members of h+*x* Persei. Data are from Humphreys (1978). Open circles are luminosity class Ia, half-filled circle class Iab, filled circle Ib and circled cross spectral class MI a-b (from Arp 1992).

## III. Conclusion

Most researchers have noticed that the observed redshifts of stars increase with the temperature of stars. In fact it is the electron number density in the atmosphere of a star increases with the temperature. The increasing number of electrons along the line of sight will lead to a larger observed redshift of the star. This is a strong support to the soft photon process proposed by author in arXiv:1305.0427 [astro-ph.HE].

## Acknowledgment

I would like to thank Dr. Nailong Wu for the correction and suggestions his made to greatly improve the English of the manuscript.